\documentclass[manuscript,screen,acmlarge,11pt]{acmart}
\usepackage{verbatim}
\usepackage{adjustbox}
\usepackage{fancyvrb}
\usepackage{algorithmic}
\usepackage{algorithm} 

\usepackage{versions}

\excludeversion{TODO}\excludeversion{MAYB} 
\excludeversion{JeremySays}
\excludeversion{CristobalSays}
\excludeversion{FabianSays}
\excludeversion{CamilaSays}
\excludeversion{INUTILE}
\excludeversion{DIMENSIONTWO}
\excludeversion{MULTIDIMENSIONAL}
\includeversion{NONANONYMOUS}\excludeversion{ANONYMOUS}
\includeversion{HEADPARAGRAPH}
\includeversion{BRIDGEPARAGRAPH}
\excludeversion{LONG}\includeversion{SHORT}
\excludeversion{VLONG}

\newtheorem{property}{Property}

\providecommand{\idtt}[1]{\ensuremath{\mathtt{#1}}}

\AtBeginDocument{%
  \providecommand\BibTeX{{%
    \normalfont B\kern-0.5em{\scshape i\kern-0.25em b}\kern-0.8em\TeX}}}

\begin{document}

\title%
[Fun Maximizing Search]%
{Fun Maximizing Search, (Non) Instance Optimality, and Video Games for  Parrots}

\author{Jérémy Barbay}
\email{jeremy@barbay.cl}
\orcid{0000-0002-3392-8353}
\affiliation{%
  \institution{
   Departamento de Ingenieria Informatica y Ciencias de la Computación (DIICC), 
    Facultad de Ingenieria (FI),
    Universidad de Concepción (UdeC)
}
  \streetaddress{Edmundo Larenas 219}
  \city{Concepción}
  \state{Region Biobio}
  \country{Chile}
}

\begin{teaserfigure}\begin{center}
\begin{minipage}[b]{.45\textwidth}
  \includegraphics[width=\textwidth]{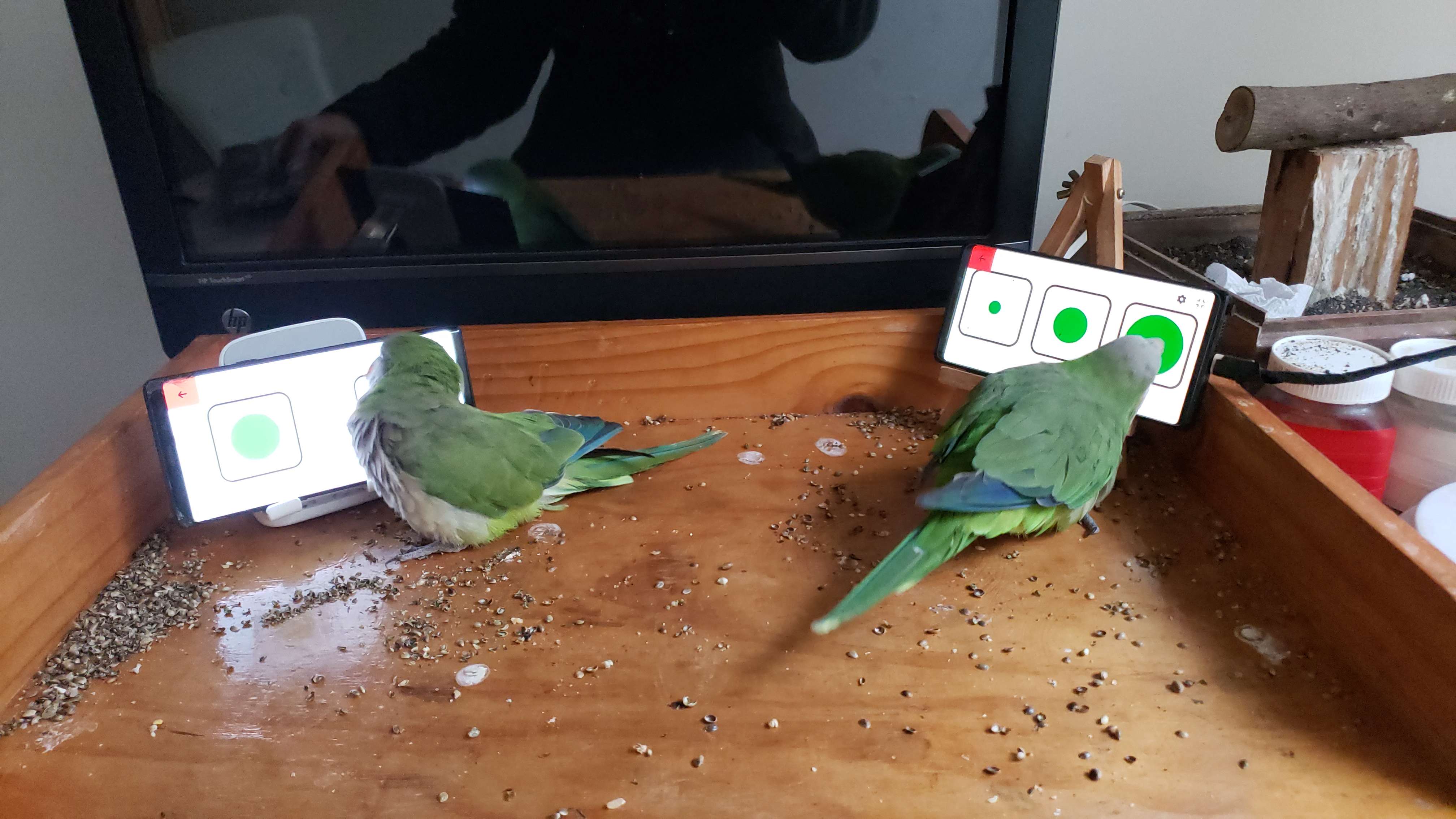}
  \caption{Two Quaker Parrots playing \texttt{InCA-ClickInOrder} on a dense array in one dimension: they have to click the values represented on the screen in decreasing order, for various representation mode of the values. In higher levels of difficulty, the values are masked by empty boxes after the first click or a parameterized amount of time, the values are displayed on a sparse grid, and/or on a grid in two dimensions. Finding a level of difficulty which is not too hard but still challenging is a difficult task, currently performed by hand.}
  \label{fig:teaser-ClickInOrder}
  \end{minipage}
\begin{minipage}[b]{.45\textwidth}
  \includegraphics[width=\textwidth]{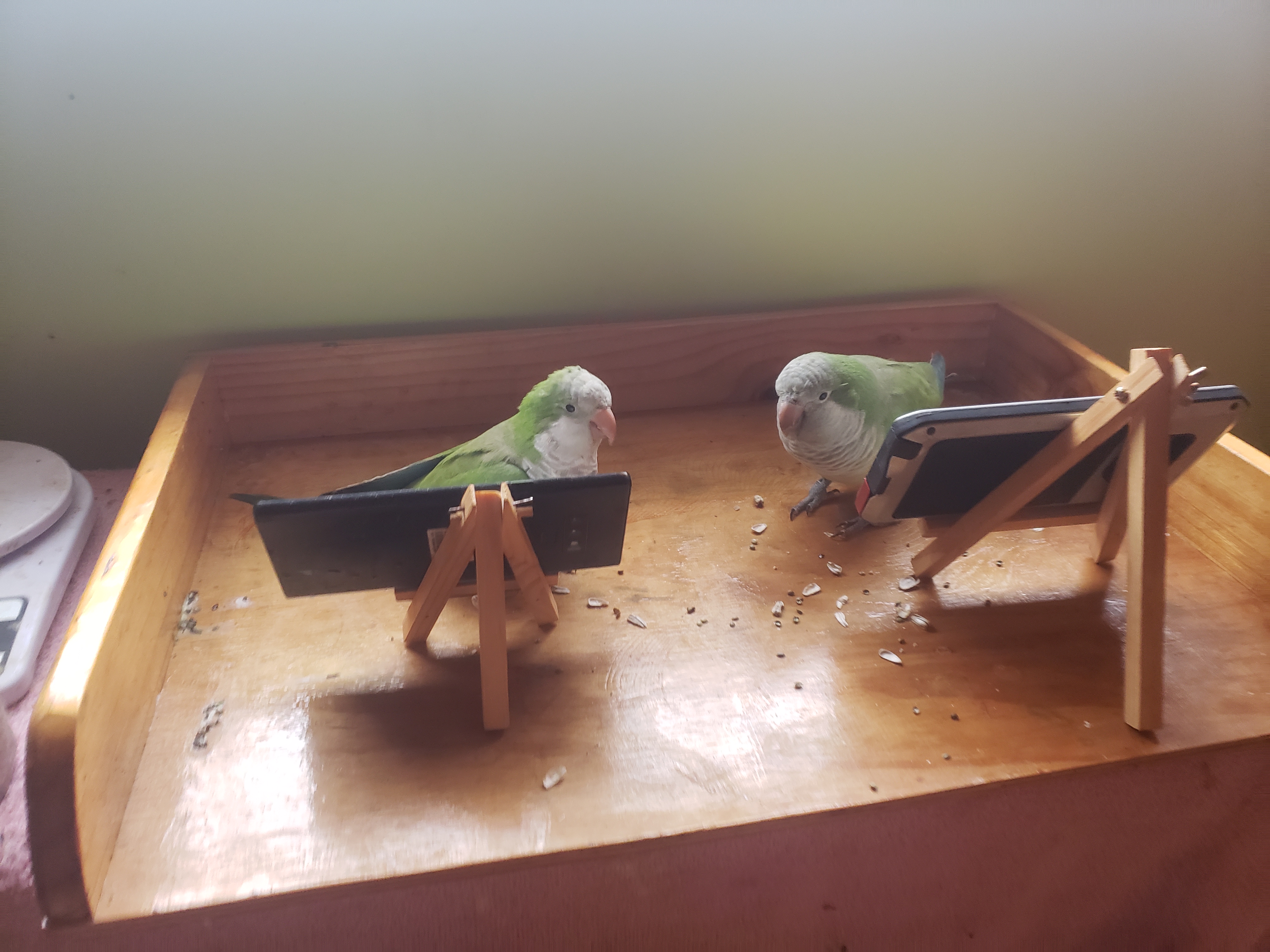}
  \caption{Two Quaker Parrots playing video games on cell phones in \emph{masked mode}, where the experimenter does not see the screen and only hears the feedback from the application indicating which type of reward to give the subjects, to insure that the subjects do not get any (voluntary or unvolutary) hints from the experimenter, hence eliminating the risk of a \emph{Clever Hans} effect~\cite{2015-CleverHans-MichaelTrestman}. }
  \label{fig:teaser-MaskedMode}
  \end{minipage}
  \end{center}
\end{teaserfigure}

\begin{abstract}
Computerized Adaptive Testing (CAT) measures an examinee's ability while adapting to their level.  Both too many questions and too many hard questions can make a test frustrating. Are there some CAT algorithms which can be proven to be theoretically better than others, and in which framework? We show that slightly extending the traditional framework yields a partial order on CAT algorithms. For uni-dimensional knowledge domains, we analyze the theoretical performance of some old and new algorithms, and we prove that  none of the algorithms presented are instance optimal, conjecturing that no instance optimal can exist for the CAT problem.
\end{abstract}


\keywords{
  Adaptive Testing,
  Instance Optimality,
  Optimality Factor,
  Sorted Search.
}

\maketitle



\paragraph{Motivation} Computerized Adaptive Testing (CAT) is a form of computer-based test that adapts to the ability level of the test's subject, in which the next item or set of items selected to be administered depends on the correctness of the subject's responses to the most recent items administered.  Even though it was originally designed for humans, we are interested in applying the concept in the context of video games for Other Animals Than Humans (OATHs), such as in video games for Dogs~\cite{2024-ACI-DogsOrNotDogs-McgrawBoscoBreyNippertEng}, Cats~\cite{2015-HFCS-PurrfectCrime-TrindadeSousaHartVieiraRodriguesFranca} or Quaker Parrots~\cite{2022-ACI-MeasuringDiscriminationOfMonkParakeetsBetweenDiscreetAndContinuousQuantitiesThroughADigitalLifeEnrichmentApplication-BarbayJanaSepulveda}. Of particular interest is the use of such algorithms in order to automatize the search for the right level of difficulty for each subject on a particular day without depending on any technical knowledge from their guardian. This is a difficult task when this ideal level of difficulty can and will evolve with time, going up when playing regularly, and going down after a long period without playing.

\paragraph{Problem}
On one hand, failing to be able to answer correctly too many questions can be frustrating for the subject, who might get discouraged and start answering questions randomly or altogether abort the test, especially if the subject is a human child or from another species than human. On the other hand, having too many questions in a test can be both inefficient and frustrating, even if many can be answered correctly by the subject.

\paragraph{Research Questions}
Which framework could/should we use to compare such CAT algorithms?
Are there some CAT algorithms which are provably better than others in such frameworks?

\paragraph{Hypothesis}
Slightly extending with an additional measure of the number of test questions answered correctly the traditional worst case and instance optimality frameworks on algorithms (which traditionally only measure the number of tests performed) yields a partial order on CAT algorithms, among which some algorithms dominate others in term of performance.

\paragraph{Proposal}
We propose to study CAT algorithms on uni-dimensional knowledge domain (i.e. where the knowledge of the subject is measured by a single position in a sequence of completely ordered knowledge items), such as for example, for a given integer $\sigma\geq 3$ the ability to compare two values chosen randomly among $[1..\sigma]$, or memorizing a permutation of $[1..\sigma]$ in 1 or 2 dimensions on a dense or sparse grid.

\paragraph{Results}
We define a new analysis framework for CAT algorithms,  which yields a partial order on CAT algorithms. We analyze in this framework the theoretical performance of various algorithms (some inspired by known sorted search algorithms, and some proposed for this work), and we prove that non of those algorithms is  instance optimal, and conjecture that no CAT algorithm can be instance optimal.

\paragraph{Outline of the Article}
After describing more formally the motivation, the problem and giving some basic algorithms as examples of solutions in Section~\ref{sec:background}, we describe in Section ~\ref{sec:theor-fram} how we extend existing theoretical frameworks to this particular context, and we define and analyze some new algorithms in such frameworks in Section~\ref{sec:new-algorithms}.  We conclude with a summary and some perspective on future work (Section~\ref{sec:conclusion}).

\section{Background} \label{sec:background}

\begin{HEADPARAGRAPH}
Even though CAT algorithms were introduced in order to improve the learning experience of humans, we are interested in their application in the field of Ethology~\cite{Ethology-Wikipedia}, in order to improve the study of the sensory and cognitive abilities of Other Animals Than Humans (OATHs). We describe some motivating examples in Section~\ref{sec:motivation} and some naive solutions in Section~\ref{sec:naive-solutions}, which performance we will later study and compare to that of new algorithms (in Section~\ref{sec:new-algorithms}).
\end{HEADPARAGRAPH}

\subsection{Motivation}\label{sec:motivation}

\begin{HEADPARAGRAPH}
Even considering that humans are but one species among others, working with Other Animals Than Humans (OATHs) present distinct challenges from working with humans. We describe some parameterized Digital Life Enrichment Activities for OATHs in Section~\ref{sec:examples-video-games} to illustrate how those differ from traditional ``video games'' for humans, and how we formalize the concept of difficulty of a given parameterization of a Digital Life enrichment activity in Section~\ref{sec:difficulty-matrices}.
\end{HEADPARAGRAPH}

\subsubsection{Examples of Video Games for Other Animals Than Humans (OATHs)}\label{sec:examples-video-games}

Life enrichment activities have proven to be a very effective way to ethically measure the sensory and cognitive abilities of various non-human species. With the advancement of technology, these activities have been modernized into what we know today as digital life enrichment applications. In 1990, the NASA/LRC Computerized Test System \cite{1990-BRMIC-TheNASALRCComputerizedTestSystem-RichardsonWashburnHopkinsSavageRumbaughRumbaugh}  was demonstrated to produce a flexible but powerful environment for the investigation of behavioral and psychological processes. This allowed rhesus monkeys (\textit{Macaca mulatta}) from different populations and locations to be tested under comparable conditions. The authors even mention that ``\emph{the animals readily started to work even when the reward was a small pellet of chow very similar in composition to the chow just removed from the cage}'', and that ``\emph{the tasks have some motivating or rewarding of their own}''. Obtaining data from different sources for the same experiment has become easier with the current technological advancement, sharing the results of a study became easier with the advent of the internet, and replicating experiments with physical to digital media becomes easier with the use of touch screens.

\paragraph{\texttt{What is More}}
Al Aïn et al.~\cite{2008-AC-TheDiscriminationOfDiscreteAndContinuousAmountsInAfricanGreyParrots-AlAinGiretGrandKreutzerBovet} tested the discrimination abilities of African Grey (\emph{Psittacus erithacus}) parrots on discrete and continuous amounts.
\begin{LONG}
More precisely, they investigated the ability of three African grey parrots to select the largest amount of food between two sets, in two types of experiments. In the first experiment type, the subjects were tested on discrete quantities via the presentation of two distinct quantities of sunflower seeds, between 1,2,3,4 and 5 seeds. In the second experiment type, the subjects were tested on continuous quantities via the presentation of two distinct quantities of parrot formula, with amounts between 0.2,0.4,0.6,0.8 and 1 ml.  For each experiment, the two amounts were presented simultaneously and were visible at the time of choice. Albeit the subjects sometimes failed to choose the largest value, they always performed above chance, their performance improving when the difference between amounts was the greatest.
Al Aïn et al.~\cite{2008-AC-TheDiscriminationOfDiscreteAndContinuousAmountsInAfricanGreyParrots-AlAinGiretGrandKreutzerBovet} gathered the data from the experiments realized in an analogous setting, performing the following statistical tests on it:
  Chi-Squared tests for possible side bias;
  binomial test in order to decide if such accuracy was substantially better than that achieved by selecting a value uniformly at random; and
  Pearson correlation coefficient between the accuracy of the subjects' answers when asked to select the maximal out of two values on one hand, and the three variables they considered like the \emph{sum} of the values for each test,
  the \emph{difference} between the two extreme values presented within a trial and
  the \emph{ratio} of continuous quantities presented, by dividing the smallest presented value by the largest one.

  \end{LONG}
Barbay et al.~\cite{2022-ACI-MeasuringDiscriminationOfMonkParakeetsBetweenDiscreetAndContinuousQuantitiesThroughADigitalLifeEnrichmentApplication-BarbayJanaSepulveda} replicated and extended  Al Aïn et al.'s study~\cite{2008-AC-TheDiscriminationOfDiscreteAndContinuousAmountsInAfricanGreyParrots-AlAinGiretGrandKreutzerBovet} via the use of a digital life enrichment application called \texttt{What is More}, used by the subjects through a touchscreen, in a way which promoted the OATH subjects' agency and allowed to gather a larger volume of data points in a smaller amount of time than in Al Aïn et al.'s original experimental protocol~\cite{2008-AC-TheDiscriminationOfDiscreteAndContinuousAmountsInAfricanGreyParrots-AlAinGiretGrandKreutzerBovet}.
The application is highly parameterized, so that the level of difficulty, or even parameters unrelated to the skills being tested, can be adjusted to the specificities of both the species (e.g. Dog, Cat, Quaker Parrot) and of the individual being tested (e.g. subject previously trained vs new subject, color blind individual, etc.).\begin{TODO} DESCRIBE all the parameters of the setting page of what is more\end{TODO}.
The task to choose and adapt such parameters requires some knowledge both of the mechanism of the application and of the experimental protocol. Letting this task fall upon the experimenter restricts who can be an experimenter (e.g. someone with a scientific training) and excludes the possibility to enroll other persons, such as technical staff in zoological parks and citizens acting as guardians of potential subjects, in citizen science projects~\cite{CitizenScience-Wikipedia}. In particular, a system automatizing the selection and evolution of such parameters a will be essential to any citizen science project based upon guardians guiding their protegees through the use of digital life enrichment applications.

\paragraph{\texttt{Click in Order}}
Inoue and Matsuzawa~\cite{2007-CB-WorkingMemoryOfNumeralsInChimpanzees-InoueMatsuzawa} investigated the working memory of chimpanzees on two-dimensional permutations.
\begin{LONG}
They conducted a series of experiments on a group of six subjects: three mothers, and offspring pairs.
\end{LONG}
With a touchscreen that showed the Arabic numerals from 1 to 9 in random positions among 40 different locations on the screen, the subjects were trained to select the correct ascending sequence of numerals.  The \emph{masking task} consists of replacing the numbers with white squares after the first numeral gets selected.
\begin{LONG}
After performing this test, the results showed that all the subjects mastered this task, even though the performance of the younger subjects was better than the one of their mothers.
\end{LONG}
Then, the \emph{limited hold memory task} was introduced to the chimpanzees; this consisted of giving a certain amount of time before covering the numbers.
\begin{LONG}
Three different hold durations were tested: 650, 430, and 210 milliseconds. At this stage, the subjects' performance was compared with human subjects: a group of 9 university students.
\end{LONG}
Overall, the results showed that the young chimpanzees performed better than both adults and human subjects\begin{LONG}, especially the subject named Ayumu, who got approximately 79\% accuracy during the test. Inoue et al.~\cite{2007-CB-WorkingMemoryOfNumeralsInChimpanzees-InoueMatsuzawa} explained this by what is known about \emph{eidetic imagery} in humans, which declines with age\end{LONG}.
Silberberg and Kearns~\cite{2009-AC-MemoryForTheOrderOfBrieflyPresentedNumeralsInHumansAsAFunctionOfPractice-SilberbergKearns} recreated the setup of the experiments held by Inoue et al.~\cite{2007-CB-WorkingMemoryOfNumeralsInChimpanzees-InoueMatsuzawa} to show that when given enough practice, humans can achieve the performance shown by the young chimpanzees.
\begin{LONG}
They first performed the \emph{masking task}, where the numbers were replaced by white squares after the first numeral was selected, giving the subjects enough time to memorize the correct sequence. This task was held with a group of students, and the results showed that the 12 untrained human subjects performed better than any ape. For this reason, the second test focused on training to get the best results in the limited hold task. 
This second task, which was carried out by the authors instead of a group student, consists of giving an amount of delay after the task begins before covering the numbers with a white square. Thus, the authors of this study started training, doing 50 trial sessions, varying the number of sessions per day from three up to ten, with different hold duration from 250 to 100 milliseconds. At the latency of 210 ms, the shortest one used in the 2007 study, both adults were able to match the accuracy level of Ayumu, the chimpanzee who performed the best in the limited hold task. To get this result the subjects took around 2500 trials throughout their sessions.
Inoue and Matsuzawa report~\cite{2007-CB-WorkingMemoryOfNumeralsInChimpanzees-InoueMatsuzawa} that humans got up to 30\% accuracy with the 210 milliseconds task, which is similar to the one achieved by the subjects before training, so Silberberg and Kearns~\cite{2009-AC-MemoryForTheOrderOfBrieflyPresentedNumeralsInHumansAsAFunctionOfPractice-SilberbergKearns} results suggest that when given enough practice, the performance of adult humans can match the one from young chimpanzees, contradicting the conclusions of Inoue et al.~\cite{2007-CB-WorkingMemoryOfNumeralsInChimpanzees-InoueMatsuzawa}.
\end{LONG}
We are currently designing, developing, and validating a simple open-source web solution called \texttt{Click In Order} (see~Figure{fig:teaser-ClickInOrder} for a picture of two Quaker parrots playing this application) to replicate and extend the software used in both studies~\cite{2007-CB-WorkingMemoryOfNumeralsInChimpanzees-InoueMatsuzawa,2009-AC-MemoryForTheOrderOfBrieflyPresentedNumeralsInHumansAsAFunctionOfPractice-SilberbergKearns}. This application can be configured to use either the same numerical representations as in the previous studies, or other numerical representations such as a number of dots, a disk of size matching the value, or a rectangle filled with a quantity of apparent liquid proportional to the value.

\begin{TODO}
ADD Screenshot of both applications.
\end{TODO}

\begin{BRIDGEPARAGRAPH}
All applications described above are parameterized, which allows the experimenter to adjust the difficulty of the application to the subject's abilities. There are various reasons for such adjustment: one objective can be to determine the limits of the subject's abilities, another one can be to maintain the application entertaining~\cite{2019-JPP-BeyondChallengeSeekingAndSkillBuilding-TseNakamuraCsikszentmihalyi} for the subjects. Without aiming for universality, we describe some ways to formalize such parameterization of the difficulty.
\end{BRIDGEPARAGRAPH}

\subsubsection{Difficulty Matrices}\label{sec:difficulty-matrices}
\begin{HEADPARAGRAPH}
As a first approach, let's consider only \emph{quantitative} parameters, such as the size of the set of values from which to select the maximum in the application \texttt{What is More}: even though \emph{qualitative} parameters (such as the display modes of values in \texttt{What is More} and \texttt{Click in Order}) do contribute to the difficulty, as a first approach they can just be seen as different ``games'', and letting the subject choose the value of such parameter can be seen as an experiment about whether this increased agency prompts an increase in motivation and interest in participating in the experiment.

For \emph{quantitative} parameters, given a specific subject to test, we formalize the concept of difficulty  of a Digital Life enrichment activity as a matrix where each entry corresponds to the probability that the subject will successfully answer questions at the corresponding level of difficulty. 
\end{HEADPARAGRAPH}

\begin{LONG}
See Table~\ref{tab:numberOfDataPoints} for an example of the quantity of data points collected in a series of experiments with the software \texttt{What is More}.
The subject was given agency when choosing the display mode. This introduced some imbalances between the number of data points collected for each display mode, yet $5,159$ data points were gathered in only two weeks, and only through voluntary participation from the subject.
\begin{table}
\centering
\begin{tabular}{c|rrrr|r}
 Set Size & Dice & Heap &  Disc & Rectangle & Total \\  \hline
 2        & 434  & 352  &  330  & 647       & 1763  \\
 3        & 344  & 134  &  178  & 503       & 1159  \\
 4        & 90   & 66   &  95   & 147       & 389   \\
 5        & 110  & 68   &  40   & 161       & 379   \\  \hline
 total    & 978  & 620  &  643  & 1458      & 3699  \\  
\end{tabular}
\caption{Number of data points collected with one single subject, separated by display modes (``Dice'', ``Heap'', ``Disc'' and ``Rectangle'') and accumulated over all display modes (``Total'').
  The imbalance between the frequencies of the display modes and between the amounts of test results for each subjects is explained by the care to support the agency of the subject: they could interrupt the session at any time, and had the option to choose the display mode at any time (which they seldom did). Figure adapted from Barbay et al.~\cite{2022-ACI-MeasuringDiscriminationOfMonkParakeetsBetweenDiscreetAndContinuousQuantitiesThroughADigitalLifeEnrichmentApplication-BarbayJanaSepulveda}.}
\label{tab:numberOfDataPoints}
\end{table}
\end{LONG}

\begin{MULTIDIMENSIONAL}
\paragraph{One Dimension}
\end{MULTIDIMENSIONAL}
The difficulty matrix is uni-dimensional when varying the values of a single parameter: this is the simplest case, and a starting point for the study of the concept of difficulty matrices. Barbay et al.~\cite{2022-ACI-MeasuringDiscriminationOfMonkParakeetsBetweenDiscreetAndContinuousQuantitiesThroughADigitalLifeEnrichmentApplication-BarbayJanaSepulveda} basically computed for each of the two subjects  $4$ such uni-dimensional matrices, one for each of the $4$ display modes for the task of selecting the maximum value out of a set of 2,3, 4 or 5 values displayed, showing that the probability of failing the test was increasing with the number of values displayed for all display modes.

\begin{LONG}
\begin{table}
\centering
 \begin{tabular}{c|cccc|c}
Session      & Dice             & Heap            & Disc             & Rectangle       & Total           \\
\hline
17,17h       & $80 (2e^{-01})$  & $71 (9e^{-02})$ & $60 (4e^{-01})$  & $77 (2e^{-02})$ & $72 (2e^{-03})$ \\
19,11h       & $80 (6e^{-03})$  & $94 (7e^{-05})$ & $81 (1e^{-06})$  & $98 (6e^{-15})$ & $89 (1e^{-23})$ \\
20,10h       & $82 (2e^{-09})$  & $84 (2e^{-09})$ & $87 (2e^{-14})$  & $87 (9e^{-13})$ & $85 (1e^{-41})$ \\
22,09h       & $60 (5e^{-01})$  & $100 (3e^{-02})$ & (no data)        & (no data)       & $80 (5e^{-02})$ \\
23,17h       & $85 (4e^{-06})$  & $92 (2e^{-10})$ & $85 (4e^{-06})$  & $82 (6e^{-10})$ & $85 (6e^{-28})$ \\
24,09h       & $100 (3e^{-02})$ & $80 (2e^{-01})$ & (no data)        & $95 (2e^{-05})$ & $93 (4e^{-07})$ \\
26,12h       & $88 (7e^{-08})$  & $88 (2e^{-06})$ & $100 (1e^{-03})$ & $80 (2e^{-01})$ & $89 (5e^{-16})$ \\
27,09h       & $86 (3e^{-05})$  & $93 (4e^{-07})$ & (no data)        & $80 (5e^{-02})$ & $88 (9e^{-12})$ \\
29,09h       & $87 (7e^{-07})$  & $90 (4e^{-06})$ & $90 (2e^{-04})$  & $92 (2e^{-10})$ & $90 (5e^{-24})$ \\
30,10h       & $77 (1e^{-04})$  & $90 (2e^{-12})$ & $92 (1e^{-11})$  & $90 (1e^{-25})$ & $88 (6e^{-49})$ \\
31,08h       & $87 (4e^{-18})$  & $86 (3e^{-05})$ & $86 (3e^{-07})$  & $88 (2e^{-27})$ & $87 (9e^{-54})$ \\
    \hline
       Total & $84 (3e^{-52})$  & $88 (1e^{-53})$ & $86 (7e^{-45})$  & $88 (1e^{-97})$ & $87 (3e^{242})$ \\
    \end{tabular}
    \caption{Finer analysis of the a subject's performance on selecting the maximal value out of two, separated by display modes (``Dice'', ``Heap'', ``Disc'' and ``Rectangle'') and accumulated over all display modes (``Total'').
      The sessions occurred during the month of July 2022 and are identified by the date $d$ and hour $h$ (e.g. the session which occurred at 17:02 on the 17th of July 2022 is identified by the tag ``17,17h''). Each entry is in the format $a(p)$ where $a$ is the accuracy reported, and $p$ is the probability of achieving such accuracy or better by selecting answers uniformly at random.
      Note how the accuracy percentages are mostly above $80\%$, and that the probability of such accuracy or a better one to be attained by selecting answers uniformly at random is smaller than $0.001$ in almost all the cases. Values taken from Barbay et al.~\cite{2022-ACI-MeasuringDiscriminationOfMonkParakeetsBetweenDiscreetAndContinuousQuantitiesThroughADigitalLifeEnrichmentApplication-BarbayJanaSepulveda}. }
\label{tab:finerAnalysisTableMaxOutOfTwoSubject1}
\end{table}
\begin{table}
\centering
\begin{tabular}{c|cccc|c}
  Session     & Dice            & Heap             & Disc             & Rect             & Total            \\ \hline
 19,15h       & $64 (2e^{-03})$ & $73 (4e^{-05})$  & $76 (1e^{-06})$  & $67 (7e^{-08})$  & $69 (9e^{-19})$  \\
 21,09h       & $84 (7e^{-06})$ & $70 (2e^{-02})$  & $84 (3e^{-07})$  & $83 (9e^{-10})$  & $82 (2e^{-21})$  \\
 22,16h       & (no data)       & $100 (4e^{-03})$ & $100 (3e^{-01})$ & $100 (4e^{-03})$ & $100 (6e^{-06})$ \\
 23,17h       & $80 (3e^{-03})$ & $90 (4e^{-04})$  & (no data)        & (no data)        & $85 (3e^{-06})$  \\
 24,09h       & $64 (2e^{-03})$ & $100 (2e^{-04})$ & $100 (4e^{-03})$ & $90 (2e^{-07})$  & $81 (1e^{-13})$  \\
 25,10h       & $71 (1e^{-17})$ & $73 (9e^{-06})$  & $80 (1e^{-15})$  & $79 (1e^{-34})$  & $76 (1e^{-68})$  \\
 27,09h       & $75 (3e^{-16})$ & $60 (3e^{-02})$  & $77 (3e^{-06})$  & $86 (3e^{-39})$  & $80 (5e^{-58})$  \\
 28,17h       & $83 (2e^{-14})$ & $66 (2e^{-04})$  & $90 (2e^{-07})$  & $90 (5e^{-22})$  & $84 (2e^{-42})$  \\ 
  \hline
        Total & $74 (1e^{-54})$ & $73 (1e^{-21})$  & $81 (1e^{-39})$  & $82 (3e^{-112})$  & $78 (6e^{221})$  \\ 
\end{tabular}
\caption{Finer analysis of a subject's performance on selecting the maximal value out of three, separated by display mode and combined.
  The average accuracy of selecting a random positions out of three is 33\%: the observed accuracy between 64\% and 90\% indicates that the subject did much better than choosing at random. Taken from Barbay et al.~\cite{2022-ACI-MeasuringDiscriminationOfMonkParakeetsBetweenDiscreetAndContinuousQuantitiesThroughADigitalLifeEnrichmentApplication-BarbayJanaSepulveda}.}
\label{tab:finerAnalysisTableMaxOutOfThree}
\end{table}
\begin{table}
\centering
\begin{tabular}{c|cccc|c}
  Session     & Dice            & Heap            & Disc            & Rect            & Total           \\ \hline
19,16h        & $70 (4e^{-03})$ & $90 (3e^{-05})$ & $65 (2e^{-04})$ & $80 (5e^{-15})$ & $76 (1e^{-23})$ \\
 21,16h       & $50 (3e^{-03})$ & $63 (1e^{-05})$ & $54 (1e^{-05})$ & $61 (8e^{-09})$ & $57 (7e^{-19})$ \\
 26,07h       & $55 (1e^{-05})$ & $61 (9e^{-05})$ & $68 (7e^{-06})$ & $72 (4e^{-09})$ & $63 (9e^{-21})$ \\
 27,17h       & $20 (8e^{-01})$ & (no data)       & (no data)       & $80 (4e^{-04})$ & $60 (4e^{-03})$ \\
   \hline
        Total & $53 (8e^{-09})$ & $66 (1e^{-12})$ & $60 (5e^{-13})$ & $71 (4e^{-32})$ & $63 (8e^{-60})$ \\ \
\end{tabular}
\caption{Finer analysis of a subject's performance on selecting the maximal value out of four, separated by display modes.
  The average accuracy of selecting a position at random out of four is 25\%.  The observed accuracies between 53\% and 72\% indicate that the subjects did not choose positions at random. Taken from Barbay et al.~\cite{2022-ACI-MeasuringDiscriminationOfMonkParakeetsBetweenDiscreetAndContinuousQuantitiesThroughADigitalLifeEnrichmentApplication-BarbayJanaSepulveda}.}
\label{tab:finerAnalysisTableMaxOutOfFour}
\end{table}
\begin{table}
\centering
\begin{tabular}{c|cccc|c}
Session & Dice            & Heap            & Disc            & Rect            & Total           \\ \hline
19,16h  & (no data)       & $0 (1e^{+00})$  & (no data)       & (no data)       & $0 (1e^{+00})$  \\
22,16h  & $57 (2e^{-07})$ & $65 (5e^{-09})$ & $60 (1e^{-05})$ & $65 (2e^{-17})$ & $62 (7e^{-35})$ \\
27,17h  & $62 (7e^{-15})$ & $56 (1e^{-05})$ & $40 (6e^{-02})$ & $68 (3e^{-22})$ & $62 (1e^{-39})$ \\
\hline
Total   & $60 (9e^{-21})$ & $58 (3e^{-12})$ & $52 (5e^{-06})$ & $67 (4e^{-38})$ & $62 (9e^{-72})$ \\ \
\end{tabular}
\caption{Finer analysis of a subject's performance on selecting the maximal value out of five, separated by display modes.
  The average accuracy of selecting a random position out of five  is 20\%.
  Most of the observed accuracies,  between 40\% and 68\%, indicate that the subject did not pick positions at random.
  The sessions with an accuracy of $0\%$ correspond to sessions where the game was interrupted after a few errors and no correct answer, either by the subject selecting the exit button (and selecting another display mode) or moving to another activity. Taken from Barbay et al.~\cite{2022-ACI-MeasuringDiscriminationOfMonkParakeetsBetweenDiscreetAndContinuousQuantitiesThroughADigitalLifeEnrichmentApplication-BarbayJanaSepulveda}.
}
\label{tab:finerAnalysisTableMaxOutOfFive}
\end{table}
\end{LONG}

Figure~\ref{fig:OneDimensionalMatricesLorenzo}  represent  the accuracy values from Barbay et al.'s experiments~\cite{2022-ACI-MeasuringDiscriminationOfMonkParakeetsBetweenDiscreetAndContinuousQuantitiesThroughADigitalLifeEnrichmentApplication-BarbayJanaSepulveda}, drawing 4 arrays (one for each of 4 display modes) of 4 values (one for each set size).
\begin{table}
\centering
\begin{tabular}{l|cccc|c}
Set size & Dice            & Heap            & Disc            & Rect            & Total           \\ \hline
$2$      & $84 (3e^{-52})$ & $88 (1e^{-53})$ & $86 (7e^{-45})$ & $88 (1e^{-97})$ & $87 (3e^{242})$ \\
$3$      & $74 (1e^{-54})$ & $73 (1e^{-21})$ & $81 (1e^{-39})$ & $82 (3e^{-112})$ & $78 (6e^{221})$ \\ 
$4$      & $53 (8e^{-09})$ & $66 (1e^{-12})$ & $60 (5e^{-13})$ & $71 (4e^{-32})$ & $63 (8e^{-60})$ \\ 
$5$      & $60 (9e^{-21})$ & $58 (3e^{-12})$ & $52 (5e^{-06})$ & $67 (4e^{-38})$ & $62 (9e^{-72})$ \\
\end{tabular}
\caption{Average accuracy values of a subject for the task of selecting the maximal value out of a set of 2,3,4 or 5 distinct values, for four distinct display modes.
  Each entry is in the format $a\%(p)$ where $a$ is the accuracy reported, and $p$ the probability to achieve such accuracy by selecting answers uniformly at random.
  Each column corresponds to a single difficulty matrix. Note how the accuracy always goes down when the difficulty (the set size) augments. Data compiled from that of  Barbay et al.~\cite{2022-ACI-MeasuringDiscriminationOfMonkParakeetsBetweenDiscreetAndContinuousQuantitiesThroughADigitalLifeEnrichmentApplication-BarbayJanaSepulveda}.}
\label{fig:OneDimensionalMatricesLorenzo}
\end{table}

\begin{MULTIDIMENSIONAL}
\paragraph{Two Dimensions} Such a matrix is two dimensional when exploring the difficulty of the test is a combination of two parameters. The number of combinations of parameter values can still be manageable only if the set of values are very limited. For instance, Barbay et al.~\cite{2022-ACI-MeasuringDiscriminationOfMonkParakeetsBetweenDiscreetAndContinuousQuantitiesThroughADigitalLifeEnrichmentApplication-BarbayJanaSepulveda} explored not only $4$ possible sizes of the set of values from which to select the maximum (from $2$ to $5$), but also $2$ sizes ($6$ and $9$) for the set of values from which the displayed values were randomly chosen ($\{1,2,3,4,5,6\}$ and $\{1,2,3,4,5,6,7,8,9\}$, respectively), resulting in $8$ pairs of parameter values to test for each of the $8$ matrices of size  $4\times 2$, for a total of $64$ experiments.
But limiting the study to  $2$ sizes for the set of values from which the displayed values were randomly chosen

\begin{TODO}
ADD one figure with the accuracy values from \cite{2022-ACI-MeasuringDiscriminationOfMonkParakeetsBetweenDiscreetAndContinuousQuantitiesThroughADigitalLifeEnrichmentApplication-BarbayJanaSepulveda}, for the two parameters
\end{TODO}

\paragraph{Multi Dimensional}
\end{MULTIDIMENSIONAL}

\begin{BRIDGEPARAGRAPH}
The values in the difficulty matrix decrease with the difficulty.  Finding the parameterization of the activity so that to insure a desired success rate (e.g. $80\%$) corresponds to finding the insertion rank of the success rate in a sorted array.  This is a classical algorithmic problem, \textsc{Sorted Search}, for which many  classical algorithmic solutions are known: we describe a few quickly in the next section. We will see (in Section~\ref{sec:naive-solutions}) that those can ask (too) many difficult questions while searching for the insertion rank, which motivated the search and study of some new algorithms.
\end{BRIDGEPARAGRAPH}
\subsection{Naive Solutions}\label{sec:naive-solutions}
\begin{HEADPARAGRAPH}
Given a (virtual) one dimension difficulty array $D$ (which values are unknown but can be computed) monotonically decreasing, and a difficulty objective $t$, we need algorithms to find the \emph{insertion rank} of $t$ in $D$. We describe the adaptation of such algorithms to our context and discuss in Section~\ref{sec:trad-analys-fram} how (in)adequate classical theoretical frameworks are to evaluate such algorithms for the application discussed in Section~\ref{sec:motivation}.
\end{HEADPARAGRAPH}

\paragraph{\texttt{Sequential Search}} is the classical search algorithm~\cite{1998-BOOK-TheArtOfComputerProgrammingVol3-Knuth} starting from the lowest index $l$ in $D$ and comparing each value of $D$ from this index with the difficulty threshold $t$ until either a position $p+1$ in $D$ is reached such that $D[p+1]\geq t$, i.e. $D[p]> t\geq D[p+1]$, or the last position $r$ of $D$ is such that $D[r]>t$.

\begin{LONG}
\begin{TODO}
The pseudo code of Sequential Search is given in Algorithm~\ref{alg:sequentialSearch}
\begin{algorithm}
  \caption{Sequential Search}
  \label{alg:sequentialSearch}
\begin{verbatim}

\end{verbatim}
\end{algorithm}
\end{TODO}
\end{LONG}

\paragraph{\texttt{Binary Search}} is another classical search algorithm~\cite{1998-BOOK-TheArtOfComputerProgrammingVol3-Knuth}. In our context, the algorithm starts by comparing $t$ with the value of lowest index $D[l]$, returning $l$ if $D[l]<t$. If not, as long as the lowest index $l$ is at least one less than the highest index $r$, it then compares the threshold with the value at the position $\lfloor {l+r}\over 2\rfloor$ and recurse in the half of the matrix which might contain values smaller than $t$, and returns the insertion point when it stops the recursion.

\begin{LONG}
\begin{TODO}
The pseudo code of Binary Search is given in Algorithm~\ref{alg:binarySearch}
\begin{algorithm}
  \caption{Binary Search}
  \label{alg:binarySearch}
\begin{verbatim}
function binary_search(A, n, T) is
    L := 0
    R := n − 1
    while L ≤ R do
        m := floor((L + R) / 2)
        if A[m] < T then
            L := m + 1
        else if A[m] > T then
            R := m − 1
        else:
            return m
    return unsuccessful
\end{verbatim}
\end{algorithm}
\end{TODO}
\end{LONG}

\paragraph{\texttt{Doubling Search}}~\cite{1976-IPL-AnAlmostOptimalAlgorithmForUnboundedSearching-BentleyYao} (also called \texttt{Exponential Search} and many other names) is an adaptive variant of \texttt{Binary Search}, which first ``gallops'' by performing a \texttt{Sequential Search} for $t$ on the sub array of $D$ which indices are powers of two. Once found the insertion rank of $t$ in this sub-array, it performs a \texttt{Binary Search} in the corresponding interval in $D$, considering all indices in this interval.

\begin{LONG}
\begin{TODO}
The pseudo code of Doubling Search is given in Algorithm~\ref{alg:doublingSearch}
\begin{algorithm}
  \caption{Gallop}
  \label{alg:gallop}
  \begin{algorithmic}
  \STATE Toto
  \end{algorithmic}
\end{algorithm}
\begin{algorithm}
  \caption{Doubling Search}
  \label{alg:doublingSearch}
\begin{verbatim}
// Returns the position of key in the array arr of length size.
template <typename T>
int exponential_search(T arr[], int size, T key)
{
    if (size == 0) {
        return NOT_FOUND;
    }
    int bound = 1;
    while (bound < size && arr[bound] < key) {
        bound *= 2;
    }
    return binary_search(arr, key, bound/2, min(bound + 1, size));
  }
\end{verbatim}
\end{algorithm}
\end{TODO}
\end{LONG}

\begin{MULTIDIMENSIONAL}
\subsubsection{Multi Dimensional Search Algorithms}\label{sec:multi-dimens-search}

\paragraph{Exhaustive search}
\paragraph{Radius search}
\end{MULTIDIMENSIONAL}

\begin{BRIDGEPARAGRAPH}
A theoretical framework permits to predict the relations between the performances of various algorithms on yet unknown instances. Typically, one considers the asymptotic worst case number of operations performed. For the application described in Section~\ref{sec:motivation}, this is not sufficient.
\end{BRIDGEPARAGRAPH}
\subsection{(Inadequacy of) Traditional Analysis Frameworks}\label{sec:trad-analys-fram}

\begin{HEADPARAGRAPH}
Given a difficulty array $D$, and a difficulty objective $t$, each algorithm must be evaluated in function not only of the number of values of the array computed (i.e. the number of questions made to the subject), but also about the number of comparisons with $t$ which result was negative (i.e. the question was too difficult for the subject).
\end{HEADPARAGRAPH}

\paragraph{\texttt{Sequential Search}}~\cite{1968-BOOK-TheArtOfComputerProgramming-Knuth} results in $p+1$ tests to the subject, among which only $1$ being too difficult. It is optimal in the number of ``negative tests'', but quite inefficient in the total number of tests.

\paragraph{\texttt{Binary Search}}~\cite{1968-BOOK-TheArtOfComputerProgramming-Knuth} results in $1+\log_2(r-l+1q)$ tests to the subjects, among which in the worst case $\log_2(r-l+1)$ are ``negative tests''. It is optimal (in the worst case over instances of fixed size $r-l$) in the overall number of tests performed, but potentially quite frustrating to the subject, all tests but one being ``too hard''.

\paragraph{\texttt{Doubling Search}}~\cite{1976-IPL-AnAlmostOptimalAlgorithmForUnboundedSearching-BentleyYao}  results 
in $1+2\log_2(p-l+1)$ tests to the subjects, among which in the worst case $1+\log_2(p-l+1)$ are ``negative tests''. It is within a constant factor of being optimal (in the worst case over instances of fixed input size $r-l$ and output $p$) in the overall number of tests performed, but potentially quite frustrating to the subject, with many tests in the last sequence being ``too hard''.

\begin{BRIDGEPARAGRAPH}
Being evaluated with two criteria, one algorithm is better than another one only if the first one is better than the second one in both criteria: it is a basic principle in computational geometry and yields a partial order on algorithms. We formalize its application to the context of CAT algorithms in the next section.
\end{BRIDGEPARAGRAPH}
\section{New Theoretical Framework}\label{sec:theor-fram}
\begin{HEADPARAGRAPH}The concept of a partial order on algorithms and data structures is far from new:
comparing algorithms in terms of both memory usage and running time yields a partial order, and 
so does comparing data structures in terms of supporting time for two or more operations (one ``sacrifices'' the supporting time of one operation in order to reduce the supporting time of others) or in terms of both space and supporting time. We describe it here shortly in the context of CAT algorithms, by defining  the frustration potentially generated in a subject by a CAT algorithm as a two dimensional point (Section~\ref{sec:frustration-measure}), and by defining terms covering the various possible relations between such points (Section~\ref{sec:comp-cat-algor}). 
 \end{HEADPARAGRAPH}

\subsection{Measure of Frustration}\label{sec:frustration-measure}

Given a CAT algorithm $A$, a subject $S$ and a knowledge threshold $t$, we merely define the frustration of the subject $S$ as a point in two dimensions, which coordinates are given by the total number of test performed, and by the number of tests answered negatively by the subject:
\providecommand{\frustration}{\ensuremath \Phi}

\begin{definition}[Measure $\frustration$ of Frustration]
Given a CAT algorithm $A$, a subject $S$ and a knowledge threshold $t$, w
the frustration $\frustration(A,S,t)$ generated by the algorithm $A$ in  the subject $S$ while aiming for the knowledge threshold $t$ is a pair composed of
\begin{itemize}
\item the total number of tests presented to the subject $S$ by the algorithm $A$  while aiming for the knowledge threshold $t$, and of
\item the number of tests answered negatively by the subject among those.
\end{itemize}
\end{definition}

The reason to explicitly focus on the frustration (rather than, say, the ``fun'') is so that the goal of finding better algorithms still corresponds to \emph{minimizing} the complexity measure, as in traditional complexity analysis.
\begin{BRIDGEPARAGRAPH}
The difference with traditional analysis is that the order between algorithms is partial.
\end{BRIDGEPARAGRAPH}
\subsection{Comparing CAT Algorithms for Fun}\label{sec:comp-cat-algor}
\begin{HEADPARAGRAPH}
We chose to define formally the way to compare algorithms in function of their fun factor rather than in term of the frustration potentially generated: 
\end{HEADPARAGRAPH}
\providecommand{\lessFun}{\ensuremath \subset}
\providecommand{\moreFun}{\ensuremath \supset}
\providecommand{\funInc}{\ensuremath \#}
\providecommand{\aLessFun}{\ensuremath \overset{\circ}{\subset}}
\providecommand{\aMoreFun}{\ensuremath \overset{\circ}{\supset}}
\providecommand{\aFunInc}{\ensuremath \overset{\circ}{\#}}
\providecommand{\negative}{\idtt{negative}}
\providecommand{\total}{\idtt{total}}
\begin{definition}[More \emph{or} Less Fun]
Given two CAT algorithms $A$ and $B$, of measures of frustration $\frustration(A)=(\negative_A,\total_A)$ and $\frustration(B)=(\negative_B,\total_B)$, i.e. where
$A$ yields  $\negative_A$ failed answers out of $\total_A$ and
$B$ yields  $\negative_B$ failed answers out of $\total_B$: 
\begin{itemize}
\item $A$ is \emph{More Fun} than $B$, noted $A\moreFun B$ if 
  $\negative_A<\negative_B$
  and
  $\total_A<total_B$;
\item $A$ is \emph{Less Fun} than $B$, noted $A\lessFun B$ if
  $\negative_A>\negative_B$ while   $\total_A=\total_B$;
  or
  $\total_A>\total_B$ while   $\negative_A=\negative_B$;
\item $A$ is \emph{Fun Non Comparable} than $B$, noted $A\funInc B$, otherwise.
\end{itemize}
\end{definition}

\begin{LONG}

\begin{BRIDGEPARAGRAPH}
\end{BRIDGEPARAGRAPH}

\subsection{CAT Worst Case and Instance Optimality} \label{sec:cat-worst-case}

\begin{definition}[CAT Worst Case]

\end{definition}

\begin{definition}[CAT Instance Optimality and Optimality Ratio]

\end{definition}

\begin{BRIDGEPARAGRAPH}
\end{BRIDGEPARAGRAPH}
\end{LONG}

\begin{LONG}
\section{Theoretical Analysis}\label{sec:theoretical-analysis}
\begin{HEADPARAGRAPH}HEAD\end{HEADPARAGRAPH}

\subsection{Old Algorithms}\label{sec:old-algorithms}

\begin{HEADPARAGRAPH}Transcribing  the informal analysis from Section~\ref{sec:naive-solutions} in the new framework yields the following properties.\end{HEADPARAGRAPH}
\subsubsection{Sequential Search}\label{sec:sequential-search}

\begin{property}
A \texttt{sequential search} yields $p-1$ correct answers and $1$ incorrect answer. More formally, \texttt{Sequential Search}'s measure of frustration in function of the knowledge threshold $p$ of the subject is
$$\frustration(\idtt{Sequential})=(1,p-1).$$
\end{property}

\subsubsection{Binary Search}\label{sec:binary-search}
\begin{property}
In the worst case, a \texttt{binary Search} yields $\log_2 n$ incorrect answers. So, \texttt{binary Search}'s measure of frustration in function of the size $n$ of the knowledge domain and of the knowledge threshold $p$ of the subject is
$$\frustration(\idtt{Binary})=(\log_2 n,1{+}\log_2 n).$$
\end{property}

\subsubsection{Doubling Search}\label{sec:doubling-search}

\begin{property}
In the worst case, a \texttt{doubling Search} yields $\log_2 p$ correct answers and $1{+}\log_2 p$  incorrect answers. So, \texttt{Doubling Search}'s measure of frustration in function of the size $n$ of the knowledge domain and of the knowledge threshold $p$ of the subject is
$$\frustration(\idtt{Doubling})=(1{+}\log_2 p,1{+}2\log_2 p).$$
\end{property}
\end{LONG}

\begin{BRIDGEPARAGRAPH}
\end{BRIDGEPARAGRAPH}
\section{New Algorithms}\label{sec:new-algorithms}

\begin{HEADPARAGRAPH}
The classical algorithms described in Section~\ref{sec:naive-solutions} are parts of a partial order over all TAC algorithms. We describe two additional algorithms, one with an interesting property concerning the number of negative tests performed (Section~\ref{sec:fun-search}), and the other one expressly designed to perform poorly in this regard (Section~\ref{sec:frustrating-search}), in such a way to allow to prove some interesting results in Section~\ref{sec:frustr-based-compl}.
\end{HEADPARAGRAPH}

\subsection{\texttt{Fun Search}}\label{sec:fun-search}

The TAC algorithm \texttt{Fun Search} (formalized in Algorithm~\ref{alg:funSearch}) is an adaptive variant of the algorithm \texttt{Doubling Search} where, instead of performing a \texttt{Binary Search} in the interval found through sequential search on a sub-sequence of the items (i.e. a ``gallop''), it performs another \texttt{Fun Search} in such interval.

\providecommand{\funSearch}{\texttt{Fun}}
\providecommand{\zeroes}{\idtt{zeroes}}
\providecommand{\ones}{\idtt{ones}}
\begin{algorithm}
  \caption{\funSearch}
  \label{alg:funSearch}
  \begin{algorithmic}
  \STATE \texttt{Gallop} from small value to find a range $[l..r]$ of position such that $l$ corresponds to a known item and $r$ to an unknown one.
  \WHILE{$r-l>1$}
  \STATE \texttt{Gallop} up from $l$ until finding a range $[l'..r']$ of position such that $l'$ corresponds to a known item and $r'$ to an unknown one
  \STATE $(l,r)\leftarrow(l',r')$
  \ENDWHILE
  \STATE \texttt{Return} $l$
  \end{algorithmic}
\end{algorithm}

This results in  an interesting property concerning the number of negative tests performed: it is adaptive to the number of ``ones'' in the binary writing of the output:
\begin{property}
Given a domain knowledge of size $n$ and a knowledge threshold $p$ for the subject, 
the CAT algorithm {\funSearch} (listed as Algorithm~\ref{alg:funSearch}) yields as many negative tests as the number $\ones$ of ones in the binary writing of the knowledge threshold $p$. More formally:
 \begin{align*}
  \frustration(
  \funSearch,
  n,p,\ones)
   =       (\ones,1+2\log_2 p)
 & \subset  (\log_2 n, 1+\log_2 n) \\
\end{align*}
\end{property}

\subsection{\texttt{Frustrating Search}}\label{sec:frustrating-search}

The TAC algorithm \texttt{Frustrating Search}  (formalized in Algorithm~\ref{alg:frustratingSearch})  is expressly designed to perform poorly  regarding the number of negative tests performed, in such a way to allow to prove some interesting results in Section~\ref{sec:frustr-based-compl}.
\providecommand{\frustratingSearch}{\texttt{Frustrating}}
\providecommand{\zeroes}{\idtt{zeroes}}
\providecommand{\ones}{\idtt{ones}}
\begin{algorithm}
  \caption{\frustratingSearch}
  \label{alg:frustratingSearch}
  \begin{algorithmic}
  \STATE \texttt{Gallop} from small value to find a range $[l..r]$ of position such that $l$ corresponds to a known item and $r$ to an unknown one.
  \WHILE{$r-l>1$}
  \STATE \texttt{Gallop} down from $r$ until finding a range $[l'..r']$ of position such that $l'$ corresponds to a known item and $r'$ to an unknown one
  \STATE $(l,r)\leftarrow(l',r')$
  \ENDWHILE
  \STATE \texttt{Return} $l$
  \end{algorithmic}
\end{algorithm}

The analysis of the frustration potentially caused by such an algorithm is similar to that of \texttt{Fun Search}:
\begin{property}
Given a domain knowledge of size $n$ and a knowledge threshold $p$ for the subject, 
the CAT algorithm {\frustratingSearch} (listed as Algorithm~\ref{alg:frustratingSearch}) yields as many negative tests as the number $\zeroes$ of zeroes in the binary writing of the knowledge threshold $p$. More formally:
\begin{align*}
\frustration(\frustratingSearch,n,p,\zeroes)
 & =       (\zeroes,1+2\log_2 p) \\
 & \subset  (\log_2 n, 1+\log_2 n) \\
\end{align*}
\end{property}

As one can sees, the TAC algorithm \texttt{Frustrating Search} performs as many tests in total than the algorithm \texttt{Fun Search}, but it yields the largest amount of negative tests when the knowledge threshold is low, which is exactly when the algorithm \texttt{Fun Search} yields the smallest amount of negative tests: hence the name.

\begin{BRIDGEPARAGRAPH}
The combination of the analysis of the CAT algorithms \texttt{Fun Search} and \texttt{Frustrating Search} yields some interesting conclusions about the frustration-based complexity of the CAT problem itself.
\end{BRIDGEPARAGRAPH}

\subsection{Frustration-based Complexity of CAT}\label{sec:frustr-based-compl}

The CAT algorithm \texttt{Fun Search} is \emph{More Fun} than \texttt{Doubling Search}, and asymptotically more so than \texttt{Binary Search}. Incidentally, so is the algorithm \texttt{Frustrating Search}!
Both are incomparable with the algorithm \texttt{Sequential Search}, which performs optimally in terms of the amount of negative tests but worst in terms of the total amount of tests.
Can anything be told from those results about the  frustration-based Complexity of the problem itself?

\begin{property}
The optimality ratio of both CAT algorithms \texttt{Fun Search} and \texttt{Frustrating Search} is within $\Omega(\lg(p-l))$: those algorithms are not instance optimal.
\end{property}
\begin{proof}
The wort case of each algorithm is the best case of the other one, and the ratio between their complexity in terms of the number of negative tests performed is maximized when the output is $0$.
\end{proof}

Can one define an instance optimal~\cite{2021-CHAPTER-FromAdaptiveAnalysisToInstanceOptimality-Barbay} CAT algorithm?
We conjecture that no such algorithm exists:
\begin{conjecture}
No CAT algorithm can be instance optimal.
\end{conjecture}

\begin{MULTIDIMENSIONAL}
\begin{BRIDGEPARAGRAPH}
\end{BRIDGEPARAGRAPH}
\section{Solutions in two dimensions}\label{sec:solut-two-dimens}
\begin{HEADPARAGRAPH}
HEAD
\end{HEADPARAGRAPH}
\end{MULTIDIMENSIONAL}

\section{Conclusion}\label{sec:conclusion}

\paragraph{Results}
Computerized Adaptive Testing (CAT) measures an examinee's ability while adapting to their level.  Both too many questions and too many hard questions can make such a sequence of tests frustrating.
We defined a framework which yields a partial order on CAT algorithms, studied CAT algorithms inspired from classical solution to the \textsc{Sorted Search} problem and defined two new CAT algorithms, which are output-adaptive in different ways, proving that none of them can be instance optimal.  A preliminary analysis seems to suggest that no CAT algorithm can be instance optimal.

\paragraph{Discussion} 
Our results are works in progress: the concepts of instance optimality and of optimality ratio in a two dimensional partial order framework on algorithm should be more formally defined.

\paragraph{Directions for Future Work}
    \begin{itemize}
  \item Implement the CAT algorithm \texttt{Fun Search}, and test it with both Human and OATHs subjects, evaluating and comparing its ability to test the subjects without frustrating them too much with that of other algorithms, such as \texttt{Binary Search}, \texttt{Doubling Search}, or even \texttt{Sequential Search} (the later might causes some ethical issues!).
  \item Define and study two dimensional Testing Domain (i.e. two dimensional difficulty matrices): the \texttt{marriage before conquest algorithm} originally defined by Kirkpatrick and Seidel~\cite{1985-SOCG-OutputSizeSensitiveAlgorithmsForFindingMaximalVectors-KirkpatrickSeidel,1986-JCom-TheUltimatePlanarConvexHullAlgorithm-KirkpatrickSeidel} to compute \textsc{Maxima sets} and \textsc{Convex Hulls} in two dimensions, later proved to be \emph{input order oblivious instance optimal} by Afshani et al.~\cite{2017-JACM-InstanceOptimalGeometricAlgorithms-AfshaniBarbayChan}, seems a good candidate:
    \begin{itemize}
  \item choose one dimension
  \item run the one dimensional algorithm (adaptive to number of ones?) at the half in this dimension
  \item run the one dimensional algorithm (adaptive to number of ones?) at the threshold in the other dimension
  \item this draws a rectangle of ``easy'' tests, and divides the space in two sub instances.
    \end{itemize}    
 Generalizing the approach to even higher dimension will have real applications too!
  \item Another extension of interest is related to iterated search algorithms (e.g. finger search trees~\cite{1998-SODA-FingerSearchTreesWithConstantInsertionTime-Brodal}): in practice, one could use the result from the same subject on the previous testing day as a starting point of the search, or a point slightly lower (proportionally to the amount of time passed since the last testing), and hopefully yield an amount of tests proportional to the change in level of the subject.
    \end{itemize}




\providecommand{\acronym}[1]{ (\textbf{#1})}
\providecommand{\student}[1]{#1}
\providecommand{\correspondingAuthor}[1]{#1}
\providecommand{\soloAuthorMark}{$^{\dagger}$}
\providecommand{\soloAuthor}[1]{#1}
\providecommand{\journalRankJCR}[2]{}
\providecommand{\journalRankSJR}[3]{}
\providecommand{\confRank}[1]{}

\bibliographystyle{ACM-Reference-Format}
\bibliography{/home/jbarbay/Evergoing/Bibliography/bibliographyDatabaseJeremyBarbay,/home/jbarbay/Evergoing/Publications/publications-ExportedFromOrgmode-Barbay}




\end{document}